\documentclass[a4paper]{article}
\usepackage{INTERSPEECH2020}
\usepackage{amsmath,graphicx}
\usepackage{hyperref}
\usepackage{siunitx}
\usepackage{tikz}
\usepackage{pgfplots}
\usepackage{booktabs}
\usepackage{multirow}
\usepackage{csvsimple}
\usepackage{flushend}

\usepgfplotslibrary{statistics}
\pgfplotsset{compat=newest}

\usepackage[colorinlistoftodos,prependcaption,textsize=tiny]{todonotes}

\def\mb{\boldsymbol}
\def\etal{et al.~}
\def\wrt{w.r.t.~}
\def\ie{i.e.~}
\def\eg{e.g.~}

\newcommand{\Fig}[1] {Fig.~\ref{#1}}

\newcommand{\Sec}[1] {Sec.~\ref{#1}}

\DeclareSIUnit\k{k}
\DeclareSIUnit\M{M}
\DeclareSIUnit\FLOP{FLOP}
\DeclareSIUnit\FLOPs{FLOPs}
\DeclareSIUnit\FLOPS{FLOPS}
\DeclareSIUnit\kFLOPs{\kilo\FLOPs}
\DeclareSIUnit\kFLOPS{\kilo\FLOPS}
\DeclareSIUnit\MFLOPs{\mega\FLOPs}
\DeclareSIUnit\MFLOPS{\mega\FLOPS}

\definecolor{col1}{RGB}{211,47,47}
\definecolor{col2}{RGB}{123,31,162}
\definecolor{col3}{RGB}{0,151,167}
\definecolor{col4}{RGB}{46,125,50}

\makeatletter
\pgfplotsset{
  grid style={dotted, gray},
  boxplot/lower notch/.initial=\pgfplotsboxplotvalue{median},
  boxplot/upper notch/.initial=\pgfplotsboxplotvalue{median},
  boxplot/notch width/.initial=0.5,
  boxplot/draw/box/.code={%
    \draw[/pgfplots/boxplot/every box/.try]
      (boxplot box cs:\pgfplotsboxplotvalue{lower quartile},0)
      -- (boxplot box cs:\pgfplotsboxplotvalue{lower notch},0)
      -- (boxplot box cs:\pgfplotsboxplotvalue{median},0.5-\pgfplotsboxplotvalue{notch width}/2)
      -- (boxplot box cs:\pgfplotsboxplotvalue{upper notch},0)
      -- (boxplot box cs:\pgfplotsboxplotvalue{upper quartile},0)
      -- (boxplot box cs:\pgfplotsboxplotvalue{upper quartile},1)
      -- (boxplot box cs:\pgfplotsboxplotvalue{upper notch},1)
      -- (boxplot box cs:\pgfplotsboxplotvalue{median},0.5+\pgfplotsboxplotvalue{notch width}/2)
      -- (boxplot box cs:\pgfplotsboxplotvalue{lower notch},1)
      -- (boxplot box cs:\pgfplotsboxplotvalue{lower quartile},1)
      -- cycle
    ;
  },
  boxplot/draw/median/.code={%
    \draw[/pgfplots/boxplot/every median/.try]
        (boxplot box cs:\pgfplotsboxplotvalue{median},0.5-\pgfplotsboxplotvalue{notch width}/2)
        --
        (boxplot box cs:\pgfplotsboxplotvalue{median},0.5+\pgfplotsboxplotvalue{notch width}/2)
    ;
  }
}

\title{CLC: Complex Linear Coding for the DNS 2020 Challenge}
\name{Hendrik Schr\"oter$^{1}$, Tobias Rosenkranz$^{2}$, Alberto N. Escalante-B.$^{2}$, Andreas Maier$^{1}$}
\address{
  $^1$ Friedrich-Alexander-Universit\"at Erlangen-N\"urnberg, Pattern Recognition Lab\\
  $^2$ Sivantos GmbH, Research and Development, Erlangen, Germany
}
\email{hendrik.m.schroeter@fau.de}

\begin{document}
\maketitle
\begin{abstract}
  Complex-valued processing brought deep learning-based speech enhancement and signal extraction to a new level.
  Typically, the noise reduction process is based on a time-frequency (TF) mask which is applied to a noisy spectrogram. Complex masks (CM) usually outperform real-valued masks due to their ability to modify the phase.
  Recent work proposed to use a complex linear combination of coefficients called complex linear coding (CLC) instead of a point-wise multiplication with a mask.
  This allows to incorporate information from previous and optionally future time steps which results in superior performance over mask-based enhancement for certain noise conditions.
  In fact, the linear combination enables to model quasi-steady properties like the spectrum within a frequency band.
  In this work, we apply CLC to the Deep Noise Suppression (DNS) challenge and propose CLC as an alternative to traditional mask-based processing, \eg used by the baseline.

  We evaluated our models using the provided test set and an additional validation set with real-world stationary and non-stationary noises. 
  Based on the published test set, we outperform the baseline \wrt the scale independent signal distortion ratio (SI-SDR) by about \SI{3}{\dB}.

\end{abstract}

\section{Introduction}
\label{sec:intro}

% Rel. work
% 
% - CLC
% 
% - Real-Time/Delay
% 
% - Normalization
% 
% - Conv vs RNN (performance / parameter complexity / FLOPs complexity)
% 
% - Loss functions

Monaural speech enhancement is an important part in many algorithms such as automatic speech recognition, video conference systems, as well as assistive listening devices.
Most state-of-the-art approaches work in the short-time Fourier transform (STFT) representation and estimate a TF mask using a deep neural network.
The estimated masks are usually well-defined and limited by an upper bound to improve stability of the network training.
However, previous work has shown that especially complex masks are rather hard to estimate directly and it is beneficial to compute the loss based on the enhanced spectrogram or time domain audio \cite{weninger2014discriminatively, erdogan2015phase, tan2019complex, roux2019phasebook}.
Weninger \etal\cite{weninger2014discriminatively} used real-valued masks and computed the loss on the enhanced and clean magnitudes denoted as magnitude signal approximation (MSA) instead of the predicted masks, denoted as mask approximation (MA).
Tan \etal\cite{tan2019complex} showed that this phenomena also holds for the complex domain.
Using complex masks, such as the complex ideal ratio mask (cIRM), the original signal can be ideally reconstructed.
That is, cIRM is theoretically able to modify the phase and rotate it back to the original clean phase.
While these masks are typically unbounded, in practice, the network output is bounded by an activation function to reduce the search space.
The authors showed that directly estimating a cIRM, \ie via complex mask approximation (CMA) performs worse compared to computing the loss based on complex spectrograms approximation (CSA) or time-domain signal approximation (SA).
Directly estimating the complex spectrograms, however, gives the network a huge degree of freedom which might also result in signal degradation for unseen noise types.

Le Roux \etal\cite{roux2019phasebook} also compared CMA, CSA, and a loss function based on the time-domain signal called waveform approximation (WA), first proposed by \cite{wang2018end}.
WA outperforms both CMA and SA, which provides evidence that even though complex masks allow to modify the phase and to reconstruct the ideal clean signal, they are hard to directly estimate.
Le Roux \etal also used a codebook representation to further reduce the search space for the neural network.
Their best model used a codebook containing \num{12} complex values.

Many algorithms process the noisy signal in an offline fashion \cite{lu2013speech, pascual2017segan, roux2019phasebook, williamson2018monaural,wang2018end,zhao2018convolutional} or introduce large delays, which is not viable for a lot of applications.
For instance, Zhao \etal\cite{zhao2018convolutional} or Le Roux \etal\cite{roux2019phasebook} used bidirectional recurrent neural networks (RNNs) or Pascual \etal\cite{pascual2017segan} used an encoder/decoder architecture with skip connections, both methods requiring the full audio signal.
Instead, the Interspeech Deep Noise Reduction (DNS) challenge \cite{DNSChallenge2020} aims for methods that perform online processing with a limited delay, which is for instance required by VoIP applications.
Specifically, the lookahead is limited to \SI{40}{\ms} with a maximum frame size of also $T=\SI{40}{\ms}$, which results in an overall algorithm delay of \SI{80}{\ms}.
The challenge provides two tracks.
The real-time track limits the processing time to $T/2$ on an Intel Core i5 quad core or equivalent, whereas the second track does not make any complexity and processing limitations.
Thus, the maximum overall latency for the real-time track is \SI{100}{\ms}, which is still considered lip-synchronously \cite{bt19981359}.

In this paper, we propose to use a method called complex linear coding \cite{schroeter2020clcnet} for noise reduction.
Instead of using a complex mask that is applied per TF-bin, we propose to use a linear combination of complex valued coefficients that are applied frequency bin-wise on the current time step as well as previous time steps.
Schröter \etal\cite{schroeter2020clcnet} motivated CLC by its ability to model quasi-static properties of speech.
That is, CLC is able to reduce noise within a frequency band, while keeping the speech components.
This is especially helpful, when there are multiple speech harmonics in one frequency band or noise and speech harmonics have very similar frequencies.
Mack \etal\cite{mack2019deep} used a similar technique, which they called deep filtering.
They showed that this complex linear combination can also be seen as a filter that is applied in the complex TF domain.
Since a filter applied to multiple TF bins, it is able to recover signal degradations like notch-filters or time-frame zeroing.
%\todo{Maybe write something about phase, especially for very noisy samples}
Instead of making full use of the latency and complexity requirements by the challenge, the proposed method focuses on minimal complexity and latency of approx.~\SI{20}{\ms}.
This, e.g., relaxes the hardware as well as transmission constraints of each participants in a VoIP setting.

The rest of the paper is structured as follows.
\Sec{sec:dataset} describes the dataset used for training and evaluation, and outlines the data mixing and augmentation process is outlined.
In \Sec{sec:baseline}, we shortly introduce the baseline system.
\Sec{sec:methods} formally defines CLC and depicts the proposed models.
In \Sec{sec:results}, we report the results on the provided test set as well as an additional validation set of both, our models and the baseline.
This is followed by a summary and conclusion in \Sec{sec:conclusion}.

% \todo{remove this?}
% Other approaches perform the enhancement in time domain.
% Luo \etal\cite{luo2019conv} used a fully convolutional network with an encoder/decoder structure to separate two signals without explicitly transforming the audio into TF domain.
% With their approach, they outperformed real-valued mask based methods.

\section{Dataset}
\label{sec:dataset}

%\noindent\textbf{DNS Training Dataset}\vspace{.2em}
\subsection{DNS Training Dataset}

\noindent The provided DNS training dataset consists of clean speech samples from the Librivox dataset \cite{panayotov2015librispeech} from \num{2150} speaker summing up to approx.~\SI{441}{\hour}.
The noise dataset contains samples from the Audioset \cite{gemmeke2017audio}, Freesound and Demand \cite{thiemann2013diverse} databases summing up to \num{70000} samples and at least \num{150} noise classes.

%\vspace{.5em}
%\noindent\textbf{DNS Test Dataset}\vspace{.2em}
\subsection{DNS Test Dataset}

\noindent The DNS test set contains overall 900 synthetic clips with reverberant and non-reverberant speech as well as real recordings without ground truth collected at Microsoft or taken from Audioset.

%\vspace{.5em}
%\noindent\textbf{Additional Training Data}\vspace{.2em}
\subsection{Additional Training Data}
\label{ssec:additional_data}

\noindent In addition to the provided datasets, we incorporated some speech and noise samples from different databases.
We used about \num{4000} samples from the EUROM database \cite{chan1995eurom} from the languages English, German, Swedish, Norwegian, Danish and French.
Furthermore, we used about \num{3000} samples from the TIMIT dataset \cite{zue1990speech}.
We extended the noise samples with \num{750} manually-selected noise samples from Audioset as well as about \num{1500} samples from the RNNoise dataset \cite{valin2018rnnoise}.
For all datasets, additional samples were excluded in a validation and test set.
Speech samples were split speaker exclusive.
The test set contains overall about \num{3000} noisy samples mixed with SNRs of $\{0, 5, 10, 20\}$.

%\vspace{.5em}
%\noindent\textbf{Mixing and Augmentation Process}\vspace{.2em}
\subsection{Mixing and Augmentation Process}
\label{ssec:augmentation}

\noindent We deployed our own signal mixing algorithm to generate noisy samples as well as ground truth signals.
The noisy mixtures were created by sampling up to four noises from the noise training set with various SNRs of $\{-5, 0, 5, 10, 20, 40\}$.
To simulate room environments, we used pyroomacoustics \cite{scheibler2018pyroomacoustics} and simulated various shoe box rooms with a max size of $[12, 8, 3.5]$ and $T_{60}$ between \num{30} and \SI{3000}{\ms}.
The generated room transfer functions (RTFs) were applied to \SI{50}{\percent} of the input speech samples.
In \Sec{sec:methods}, we present two models that are trained with slightly different clean speech targets.
One model uses the reverberant speech as target, the other model uses the close-source recordings before applying the RTFs.
To ensure that the alignment of clean and noisy is the same in the latter case, the non reverberant speech samples were delayed until the peak of the RTF.
Additionay, we randomly applied a gain change of $\{-6,0,6\}$ \si{\dB}.

%\vspace{.5em}
%\noindent\textbf{Preprocessing and Normalization}\vspace{.2em}
\subsection{Preprocessing and Normalization}
\label{ssec:preprocessing}

\noindent We process the noisy input data using a standard STFT with a \SI{20}{\ms}
Hamming window which corresponds to \num{161} frequency bins and \SI{75}{\percent} overlap resulting in a frequency resolution of \SI{100}{\Hz} per frequency bin.
This frequency resolution is slightly smaller then the baseline \cite{xia2020weighted} and results in a shorter delay.
However, it is a lot higher when compared to the original CLCNet \cite{schroeter2020clcnet}.

Since our input and output is complex valued, we cannot use log-power spectra with mean/variance normalization like \cite{xia2020weighted}.
Instead, similar to \cite{schroeter2020clcnet}, we normalize the complex spectrum using unit norm:
\begin{equation}
  \mb{X}_{\text{norm}}[t, f] = \frac{\mb{X}[t, f]}{\mb\hat{\mb\mu}[k, f]}\text{ ,}
\end{equation}
where $\mb X$ is the complex spectrum, $\mb\hat{\mb\mu}$ a mean estimate of $|\mb X|$, and $t$ and $f$ are time and frequency bins.
This is done in an online fashion, \ie $\mb\hat{\mb\mu}$ is estimated as follows:
\begin{equation}
  \mb\hat{\mb\mu}[t, f] = \alpha \mb\hat{\mb\mu}[t - 1, f] + (1 - \alpha) |\mb X|[t, f]  \text{\ .}
\end{equation}
The decaying factor $\alpha$ was set to \num{0.99}.
The unit normalization enhances the magnitude of the weaker parts in the spectrum while not modifying the noisy phase.

\section{Baseline}
\label{sec:baseline}

% Describe baseline, i.e. real-valued mask based approach, network structure, loss etc

Xia \etal \cite{xia2020weighted} used a simple GRU based architecture, which is similar to what we propose.
The input is transformed into TF domain using an STFT with a window size corresponding to \SI{32}{\ms}.
After applying \si{\dB}-scaling, it is mean/variance normalized using an exponential decay in an online fashion.
They used a 3 layer GRU followed by a fully connected (FC) layer and sigmoid activation to predict a real valued mask.
The network overall contained \SI{1.26}{M} parameters.
Their main contribution was a weighted SDR based loss function.

\section{Methods}
\label{sec:methods}

\subsection{Complex Linear Coding}
\label{ssec:clc}

\begin{figure}[htb]
  \centering
  \includegraphics[width=\linewidth]{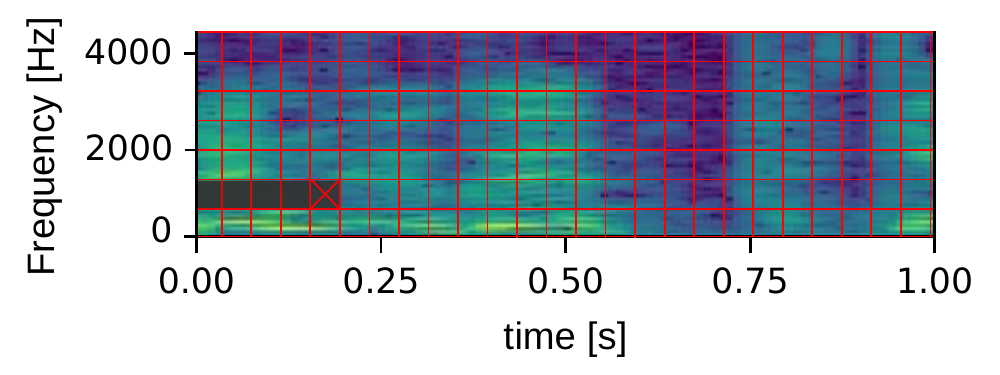}
  \caption{
    Schematic figure of complex linear coding.
    The red grid represents the TF bins, the gray boxes represent exemplary CLC coefficients of order $N=5$.
    The output TF computed by the linear combination of Eq.~\ref{eq:CLC} is marked with a red cross.
    Note that the size of the TF bins is not true to scale.
  }
  \label{fig:clc_schematic}
\end{figure}

CLC was introduced in the context of hearing aids \cite{schroeter2020clcnet}.
Motivated by linear predictive coding (LPC), CLC is able to model periodic properties within a frequency band.
Especially when dealing with wideband spectrograms with a poor frequency resolution, CLC outperforms standard real or complex mask based methods.
Using wideband spectrograms is often necessary in low-latency settings, since a narrowband spectrogram with a higher frequency resolution requires larger processing windows.
In wideband spectrograms, speech harmonics may not be clearly separated so multiple harmonics can lie within a single frequency band.
This results in cancellation due to multiple frequencies being superimposed within that band.
The periodic structure can be modeled by CLC. A schematic figure of CLC is shown in \Fig{fig:clc_schematic}.

\noindent Formally, complex linear coding is defined as
\begin{equation}
  \vspace{-0.4em}
  \mb{\hat{S}}(k, f) = \sum_{i=0}^{N} \mb{A}(k, i, f) \cdot \mb{X}(k - i + l, f)\text{\ ,}
  \label{eq:CLC}
\end{equation}
\vspace{.5em}
where $\mb{A}$ are the complex coefficients, $\mb{\hat{S}}$ the enhanced spectrogram and $N$ the CLC order.
$l$ is an optional offset parameter, which allows to incorporate future context in the linear combination when $l \ge 1$.
Theoretically, $l$ can also be negative which results in a prediction of the $l$-th frame in the future.
For $l=-1$, the linear combination is equivalent to the one in LPC.
In all experiments, we chose $N=5$ and $l=0$.

\subsection{Network Architecture}

\begin{figure}[hbt]
  \centering
  \includegraphics[width=\columnwidth, trim=0.2cm 6.5cm 0.5cm 0.1cm, clip]{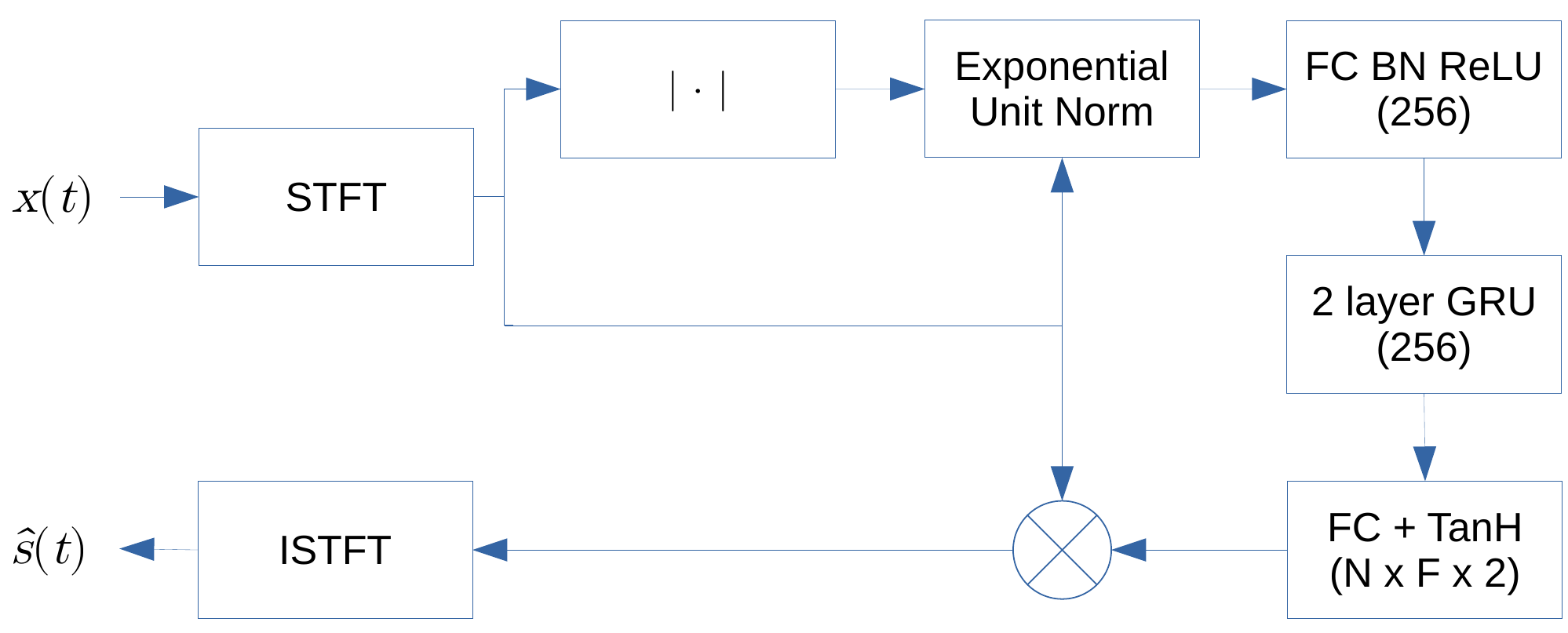}
  \caption{Flow chart of the proposed architecture. The fully connected output layer has the size $N\cdot F\cdot 2$, where $N$ is the CLC order, $F$ the frequency bins and $2$ for the real and imaginary part of the complex number. $\otimes$ denotes the complex linear combination of equation \ref{eq:CLC}.}
  \label{fig:flowchart}
\end{figure}

We used a simple network architecture similar to the baseline.
Instead of using a 3 layer GRU, we used an input layer with a fully connected layer, batch normalization and a ReLU activation.
The majority of parameters is in the output layer, since it produces a $N=5$ complex valued coefficients per frequency bin.
We used a tanh activation function for the output layer, since we need an output range from \numrange{-1}{1} for the complex coefficients.
The CLC network flow chart is shown in \Fig{fig:flowchart}.

We trained the network using PyTorch \cite{paszke2017pytorch} for \num{200} epochs with an initial learning rate of \num{0.001} and a batch size of 32.
For optimization, we used AdamW \cite{loshchilov2019decoupled} with a weight decay of \num{1e-7} and gradient clipping of \num{0.25}.
As a loss function, standard mean squared error on the time domain signal was used. We found that this outperforms a loss computed on the complex spectrogram, which are similar findings as by \eg \cite{roux2019phasebook, schroeter2020clcnet}.

We provide an open source PyTorch module including model weights and a script to process noisy input files based on PyTorch JIT\footnote{\url{https://github.com/Rikorose/clc-dns-challenge-2020}}.

% - CLC
% 
% - Network structure + graphic
% 
% - Complexity + Runtime (+ maybe comparison with baseline)
% 
% - Deverberation vs standard processing (Opensourcing models?)

\section{Results}
\label{sec:results}

%- Results on noreverb test set
%
%- Results on withreverb test set
%
%- Example image of noisy/baseline/clc-standard/clc-deverb

\noindent This section describes the qualitative results based on the provided test set of our two models.

\vspace{.5em}
\noindent\textbf{DNS Test Set}\vspace{.2em}

\noindent As explained in \Sec{ssec:augmentation}, we trained the first model on the reverberant clean target and thus only denoises its input, whereas the second model was trained to also deverberate the input signal.
Since the provided clean targets of the test set were also reverberant, we submitted the former.
As objective metrics, we use the scale independent signal distortion ratio (SI-SDR) \cite{roux2019sdr}, the short-time objective intelligibility (STOI) \cite{taal2011stoi}, and the RMSE on the time domain audio signal.
Table \ref{tab:test_results} shows the results based on the published test.
\begin{table}[hb]
  \caption{
    Objective results based on the DNS test sets.
  }
  \label{tab:test_results}
  \centering
  \resizebox{\linewidth}{!} {
  \robustify\bfseries
  \sisetup{
    table-number-alignment = center,
    table-figures-integer  = 1,
    table-figures-decimal  = 3,
    table-auto-round = true,
    detect-weight = true
  }
  \begin{tabular}{l S[table-figures-decimal=2] S S S[table-figures-decimal=2] S S}
    \toprule
     & \multicolumn{3}{c}{Non-Reverb. Test Set} &  \multicolumn{3}{c}{Reverb. Test Set}\\
    Model & \multicolumn{1}{c}{SI-SDR} & \multicolumn{1}{c}{STOI} & \multicolumn{1}{c}{RMSE} & \multicolumn{1}{c}{SI-SDR} & \multicolumn{1}{c}{STOI} & \multicolumn{1}{c}{RMSE} \\
    \midrule
    Noisy & 9.070872555549721 & 0.902586305141449 & 0.020656994256811837 & 9.03278336063687 & 0.847021206219991 & 0.0206569942071413 \\
    Baseline & 12.466973514323266 & 0.8885862731991006 & 0.013119599376458813 & 9.180369998202481 & 0.8127323526866549 & 0.018530344311977 \\
    CLC & 15.436906592051189 & 0.9309288303057353 & 0.011548545211553574 & 12.57501075108846 & 0.868169382015864 & 0.015980440098792313 \\
    CLC$_{\text{devb}}$ & 15.430992358525595 & 0.9312670604387919 & 0.011662580352276564 & 1.7966242129107317 & 0.7259373768170675 & 0.04324295371770859 \\
    \bottomrule
  \end{tabular}
  }
\end{table}

\noindent For the standard CLC model, we can see a clear improvement over the noisy input, while the baseline has negative delta STOI values.
Our CLC methods outperforms the baseline by about \SI{3}{\dB} \wrt SI-SDR for the non-reverberant set and for about \SI{3.4}{\dB} for the reverberant set.
The CLC$_{\text{devb}}$ model performs worse \wrt the objective metrics on the reverberant set due to the fact that the clean targets are also reverberant.

\begin{figure}[tbh]
  \centering
  \includegraphics[width=\linewidth, trim=0 0 0 .2cm, clip]{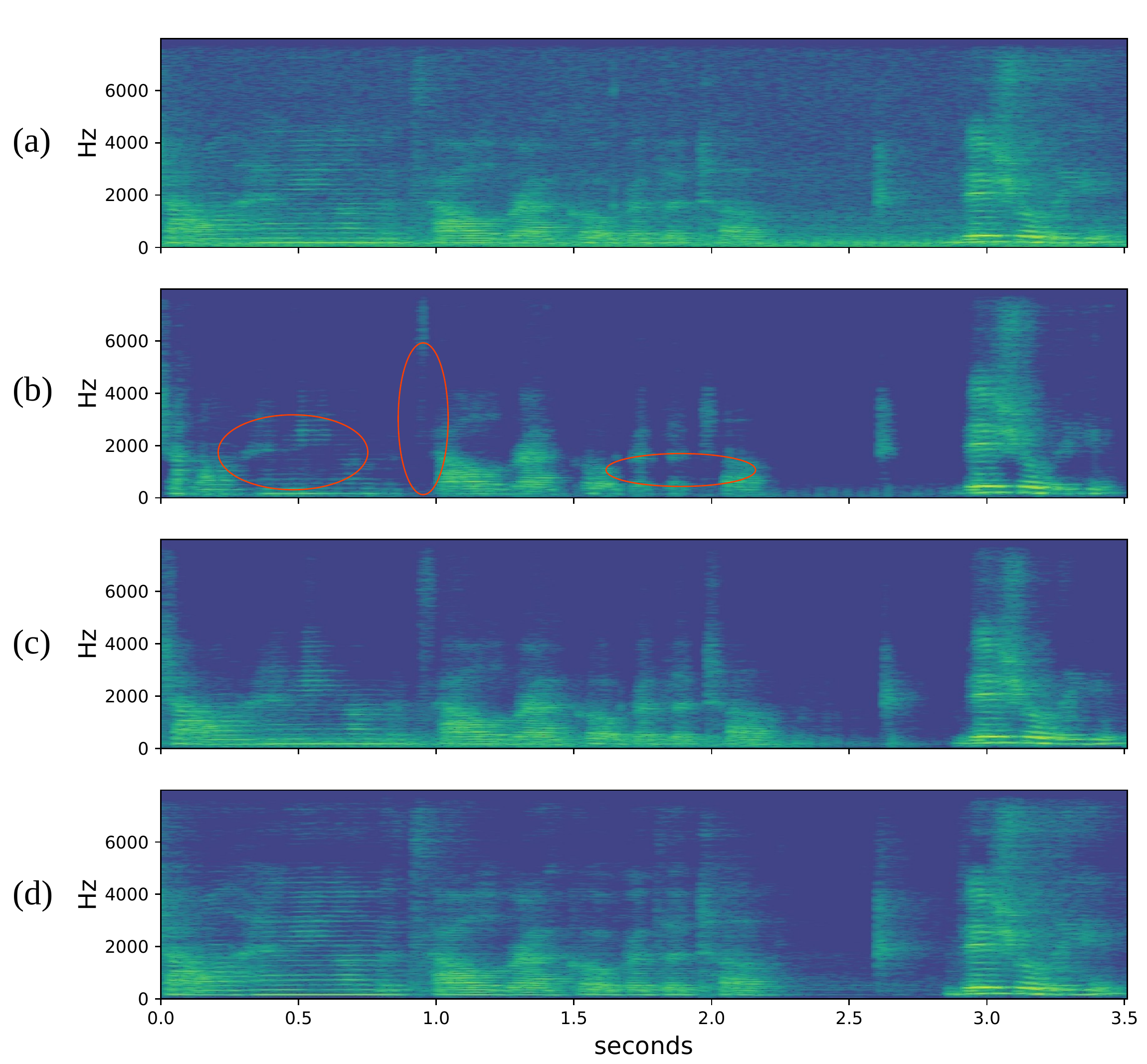}
  \caption{
    Example from the synthetic reverberant test set with noisy signal (a), baseline (b), CLC (c) and clean (d).
    Especially for reverberant signals, the baseline seems to also suppress parts of voiced and unvoiced speech.
    As a result, the intelligibility suffers and the enhanced signal is not pleasant to listen to.
  }
  \label{fig:dns_test_example}
\end{figure}

Compared to the baseline, CLC is very robust and does not degrade the speech signal to a high degree.
\Fig{fig:dns_test_example} shows an example from the test set.
Here we can see that, while the baseline degrades the speech signal quite a bit, CLC is able to preserve most of the voiced and unvoiced parts.
For transient noises like keyboard typing however, the baseline performs slightly better.
This may be due to the property of CLC, to model the more long-term, quasi stationary parts of speech and noise.

\vspace{.0em}
\noindent\textbf{DNS Blind Test Set}\vspace{.2em}

\noindent Based on the blind test set, we can also see that CLC outperforms the baseline by a delta mean opinion score (dMOS) of \num{0.17}.
Furthermore, CLC performs better on reverberant and real world data, while the baseline only performs well on the synthesized no-reverb.\ data.

\vspace{.5em}
\noindent\textbf{Our Test Set}\vspace{.2em}

\noindent Additionally to results on the published DNS test sets, we provide results based on our test set as described in \Sec{ssec:additional_data}.
As shown in Tab.~\ref{tab:test_results_own}, CLC outperforms the baseline by a large amount.
The baseline again results in a deterioration of the STOI metric.

\begin{table}[hb]
  \caption{
    Objective results based on our test set.
  }
  \label{tab:test_results_own}
  \centering
  % \resizebox{\linewidth}{!} {
  \robustify\bfseries
  \sisetup{
    table-number-alignment = center,
    table-figures-integer  = 1,
    table-figures-decimal  = 3,
    table-auto-round = true,
    detect-weight = true
  }
  \begin{tabular}{l S[table-figures-decimal=2] S S[table-figures-decimal=4] }
    \toprule
    Model & \multicolumn{1}{c}{SI-SDR} & \multicolumn{1}{c}{STOI} & \multicolumn{1}{c}{RMSE}\\
    \midrule
    Noisy & 9.0226 & 0.89904 & 0.011414 \\
    Baseline & 13.0533 & 0.873114 & 0.0060789 \\
    CLC & 17.7119 & 0.90420 & 0.00383 \\
    CLC$_{\text{devb}}$ & 16.76654 & 0.90471 & 0.00425 \\
    \bottomrule
  \end{tabular}
  % }
\end{table}

\vspace{1em}
\noindent\textbf{Complexity}\vspace{.3em}

\noindent CLC and the baseline have very similar complexities.
The DNS baseline has \SI{1.24}{M} parameters and runs in average \SI{0.6}{\ms} per \SI{32}{\ms} frame on a Intel Core i5 clocked at \SI{1.6}{\GHz}.
Note, that this is slightly higher than reported by the authors (\SI{0.16}{\ms}) on a different CPU.
Our CLC based model has about \SI{1.4}{M} parameters and runs in \SI{1.0}{\ms} per \SI{20}{\ms} frame.
While our model has a slightly higher complexity, we argue that our model performs better on real-world and on reverberant data according to the blind test set MOS results, which is highly relevant for real-world applications like VoIP scenarios.
Furthermore, the computation delay is negligible compared to the algorithm delay, and our algorithm delay is \SI{8}{\ms} less than the baseline.

\section{Conclusion}
\label{sec:conclusion}

In this paper, we presented a method based on complex linear coding.
We have shown, that CLC is very robust in a variety of speech and noise conditions and thus outperforms the baseline.
Especially for very noisy SNRs with quasi-static noisy, CLC outperforms real- and complex-valued mask-based methods.
For transient noises like \eg keyboard typing, however, CLC seems not to be able to adopt the noise fast enough.
Also, the challenge requirements \wrt processing time and introduced latency require at least desktop hardware and are not suitable for mobile or embedded devices.
Since this also allows large processing windows, resulting in a high frequency resolution, a complex mask is probably sufficient.
For smaller processing windows ({\small$\ll\SI{10}{\ms}$}), the originally proposed CLCNet has shown to outperform mask based processing.

\bibliographystyle{IEEEtran}
\bibliography{refs}

\end{document}